# Jump relations across a shock in non-ideal gas flow

## R. K. Anand
### Department of Physics, University of Allahabad, Allahabad-211002, India
E-mail: anand.rajkumar@rediffmail.com

**Abstract** Generalized forms of jump relations are obtained for one dimensional shock waves propagating in a non-ideal gas which reduce to Rankine-Hugoniot conditions for shocks in idea gas when non-idealness parameter becomes zero. The equation of state for non-ideal gas is considered as given by Landau and Lifshitz. The jump relations for pressure, density, temperature, particle velocity, and change in entropy across the shock are derived in terms of upstream Mach number. Finally, the useful forms of the shock jump relations for weak and strong shocks, respectively, are obtained in terms of the non-idealness parameter. It is observed that the shock waves may arise in flow of real fluids where upstream Mach number is less than unity.

## 1. Introduction

Shock phenomena such as a global shock resulting from a stellar pulsation or supernova explosion passing outward through a stellar envelope, or perhaps a shock emanating from a point source such as a man-made explosion in the Earth's atmosphere or an impulsive flare in the Sun's atmosphere, are of tremendous importance in astrophysics and space sciences. Shock waves are common in the interstellar medium because of a great variety of supersonic motions and energetic events, such as cloud-cloud collision, bipolar outflow from young protostellar objects, powerful mass loss by massive stars in a late stage of their evolution (stellar winds), supernova explosions, central part of star burst galaxies, etc. Shock waves are also associated with spiral density waves, radio galaxies, and quasars. Similar phenomena also occur in laboratory situations, for example, when a piston is driven rapidly into a tube of gas (a shock tube), when a projectile or aircraft moves supersonically through the atmosphere, in the blast wave produced by a strong explosion, or when rapidly flowing gas encounters a constriction in a flow channel or runs into a wall.

Rankine (1870) published his dissertation on shock waves in 1870. Hugoniot's work (Hugoniot, 1889) was published in 1889. Their equations for shock jumps in particle velocity, stress, and specific internal energy have become known as the Rankine-Hugoniot conditions. These conditions are derived from the conservation laws of mass, momentum and energy under the assumption that the shock is a single unsteady wave front with no thickness or a single steady wave of finite thickness. The study of shocks generated due to explosions or implosions in an ideal gas has received much attention in the past decades and the mainstay of theoretical description is still the Rankine-Hugoniot conditions. Many theoretical and experimental studies were reported by various investigators on planer, cylindrical and spherical shock waves since the pioneering work of Guderlay (1942), Taylor (1950a, 1950b), Sedov (1959), and von Neumann (1947) on strong shock waves from point-source (mass-less) explosions. Shock waves of moderate strength emanating from point sources were theoretically studied by Sakurai (1953, 1954, 1964), Oshima (1960), Freeman (1968), and numerically by Goldstine and von Neumann (1955), and Bach and Lee (1970). In the other extreme of the nearly linear regime, when the shock has become rather weak Whitham (1960) proposed a general theory to cover such flows. A number of approaches, including similarity method (Sedov, 1959; Zel'dovich and Raizer, 1966), power series solution method (Sakurai, 1953, 1954, 1964; Oshima, 1960; Freeman, 1968), CCW method (Chester, 1954; Chisnell, 1955; Whitham, 1958), modified CCW method (Yousaf, 1987, 1988) have been used for theoretical investigations of shock waves in different media. Rankine-Hugoniot shock conditions play a crucial role in all the above well known methods. Sanjiva (1992), Gavrilyuk and Saurel (2007), Baty et al. (2008), Emanuel (2010), Kjellander et al. (2010) and other authors have studied shock jump relations in homogeneous and inhomogeneous media.

The study of shock waves through a non-ideal gas is of great technical interest in many industrial applications such as chemical, nuclear and aerospace. Shock waves are generated by point explosions (nuclear explosions and detonation of solid explosives, solid and liquid propellants rocket motors), high pressure gas containers (chemical explosions) and laser beam focusing. Shock wave problems also arise in astrophysics, hypersonic aerodynamics and hypervelocity impact. An understanding of the properties of the shock waves both in the near-field and the far-field is useful with regard to the characteristics such as shock strength, shock overpressure, shock speed, and impulse. This has developed my interest in obtaining the jump relations for weak and strong shocks in non-ideal gas considering equation of state given by Landau and Lifshitz (1958).

In the present paper, generalized forms of jump relations for one-dimensional shock waves propagating in a non-ideal gas are derived which reduce to Rankine-Hugoniot conditions for ideal gas when non-idealness parameter becomes zero. The jump relations for pressure, density, temperature, and particle velocity are obtained, respectively in terms of upstream Mach number. The expression for change in entropy across the shock is also obtained in terms of the upstream Mach number. Further, the shock jump relations for pressure, density, and particle velocity in terms of non-idealness parameter are obtained for the two situations: viz., (i) when the shock is weak and (ii) when it is strong, simultaneously. These handy forms of jump relations for various flow quantities are very useful in the theoretical and experimental investigations of the weak and strong shock waves in real fluids. In the present theoretical investigation it is found that the shock waves may arise in flow of real fluids where upstream Mach number is less than unity i.e., $M \leq 0.5$ (approx). The rest of the paper is organized as follows: The equation of state for non-ideal gas is described in Section 2. The generalized jump relations between the upstream and downstream quantities are formulated in Section 3. Results and applications of generalized shock jump relations are discussed in Section 4. The findings are concluded in Section 5.

## 2. Equation of state

The equation of state for a non-ideal gas is obtained by considering an expansion of the pressure $p$ in powers of the density $\rho$ as (*see*, Landau and Lifshitz, 1958)

$$p = \Gamma \rho T [1 + \rho C_1(T) + \rho^2 C_2(T) + .....],$$

where $\Gamma$ is the gas constant , $p$, $\rho$ and $T$ are the pressure, density, and temperature of the non-ideal gas, respectively, and $C_1(T)$, $C_2(T)$, are virial coefficients. The first term in the expansion corresponds to an ideal gas. The second term is obtained by taking into account the interaction between pairs of molecules, and subsequent terms must involve the interactions between the groups of three, four, etc. molecules. In the high temperature range the coefficients $C_1(T)$ and $C_2(T)$ tend to constant values equal to $b$ and $(5/8)b^2$, respectively. For gases $b\rho \ll 1$, $b$ being the internal volume of the molecules, and therefore it is sufficient to consider the equation of state in the form (Anisimov and Spiner, 1972; Singh and Singh, 1998; Ojha, 2002)

$$p = \Gamma \rho T[1 + b\rho] \tag{1}$$

In this equation the correction to pressure is missing due to neglect of second and higher powers of $b\rho$, i.e., due to neglect of interactions between groups of three, four, etc. molecules of the gas. Roberts and Wu (1996, 2003) have used an equivalent equation of state to study the shock theory of sonoluminescence. The internal energy $e$ per unit mass of the non-ideal gas is given by Singh and Singh (1998) and Ojha (2002), as

$$e = p/\rho(\gamma - 1)(1 + b\rho), \tag{2}$$

where $\gamma$ is the adiabatic index. Equation (2) implies that

$$C_p - C_v = \Gamma(1 + b^2\rho^2/(1 + 2b\rho)) \cong \Gamma$$

neglecting the second and higher powers of $b\rho$. Here $C_p$ and $C_v$ are the specific heats of the gas at constant pressure and constant volume, respectively. Real gas effects can be expressed in the fundamental equations according to Chandrasekhar (1939), by two thermodynamical variables, namely by the sound velocity factor (the isentropic exponent) $\Gamma^*$ and a factor $K$, which contains internal energy as follows,

$$\Gamma^* = (\partial \ln p/\partial \ln \rho)_S \text{ and } K = -\rho(\partial e/\partial \rho)_P / p$$

Using the first law of thermodynamics and the equations (1) and (2), we obtain $\Gamma^* = \gamma(1 + 2b\rho)/(1 + b\rho)$ and $K = 1/(\gamma - 1)$, neglecting the second and higher powers of $b\rho$. This shows that the isentropic exponent $\Gamma^*$ is non-constant in the shocked gas, but the factor $K$ is constant for the simplified equation of state for non-ideal gas in the form (1).

The isentropic velocity of sound, $a$ in non-ideal gas is given by

$$a^2 = \Gamma^* p/\rho$$

## 3. Formulation of jump relations in terms of Mach number

A shock wave is not an infinitely thin discontinuity, since changes in the fluid properties always require a finite scale comparable to the mean free path of the fluid particles. The thickness is determined by the viscosity within the shock (the viscous force determines the rate of acceleration of gas through the shock, and viscous dissipation contributes to the gas heating) and by the thermal conductivity, which determines the heat flow across the shock. For most purposes, however, a shock can be idealized as a discontinuity, and the stationary flow conditions behind the shock (downstream) can be related to the conditions ahead (upstream) by requiring continuity of the mass, momentum, and energy fluxes through the shock.

When disturbances of finite amplitude are propagated in gases (destitute of viscosity or heat conductivity) discontinuities in pressure, density, velocity, and temperature of the medium may occur. These are called shock waves (or shocks) or blast waves (intense shock waves). Shock waves and the flow field become planer, cylindrically symmetric, or spherically symmetric, respectively if the energy source is in a plane, along a line, or a point in space. The appearance of shocks is a consequence of nonlinear character of the equations governing the propagation of finite disturbances that is in part responsible for the formation of discontinuities. The jump conditions at the shock are given by the principles of conservation of mass, momentum, and energy across the shock, namely,

$$\rho(U - u) = \rho_o U$$
$$p + \rho(U - u)^2 = p_o + \rho_o U^2 \qquad\qquad (3)$$
$$(U - u)^2 / 2 + p / \rho(\gamma - 1)(1 + b\rho) + p / \rho = U^2 / 2 + p_o / \rho_o(\gamma - 1)(1 + b\rho_o) + p_o / \rho_o$$

where $U$ and $u$ are shock velocity and particle velocity, respectively and the quantities with the suffix $o$ and without suffix denote values of the quantities immediately ahead of, and immediately behind, the shock front, respectively. The upstream Mach number $M_{as} \approx M$ (say), that characterizes the strength of a shock is defined as $M = U / a_o$, where $a_o \left(= \sqrt{\gamma\,\delta\,p_o / \rho_o}\right)$ is the speed of sound in the medium ahead of the shock front, $\delta = (1 + 2b\rho_o) / (1 + b\rho_o)$ and $b\rho_o$ is the non-idealness parameter. Generally, the upstream Mach number $M$ in a given problem is known and it is desired to determine the downstream Mach number $M_{bs}$. The expression for downstream Mach number $M_{bs}$ can be easily obtained in terms of upstream Mach number $M$ and written as

$$M_{bs}^2 = \frac{2 + (\gamma - 1)M^2}{2\,\gamma\,\delta(\gamma\delta - \gamma + 1)M^2 - (\gamma - 1)} \qquad\qquad (4)$$

The ratio of the different flow variables such as pressure, density, temperature, etc. across a shock wave in an ideal gas are expressed as functions of upstream Mach number $M$. They are generally referred to as the Rankine-Hugoniot jump relations or conditions. The set of equations (3) is identical with the ideal gas analogs, and the jump relations across the shock propagating in a non-ideal gas may be easily expressed as

$$p = \frac{\rho_o a_o^2}{\delta}\left\{\frac{2[\gamma(\delta^2 - 1) + \delta]M^2}{\gamma + 1} - \frac{\gamma - 1}{\gamma(\gamma + 1)}\right\}$$

$$\rho = \frac{\rho_o[\gamma^2(4\delta^2 - 3) + 2\gamma(2\delta - 1) + 1]M^2}{(\gamma + 1)[(\gamma - 1)M^2 + 2]}$$

$$T = \frac{T_o[2 + (\gamma - 1)M^2]\{2\gamma[\gamma(\delta^2 - 1) + \delta]M^2 - (\gamma - 1)\}}{[4\gamma\,\delta(\gamma\delta - \gamma + 1) + (\gamma + 1)^2]M^2} \qquad (5)$$

$$u = \frac{2a_o(\gamma + 1)}{[\gamma^2(4\delta^2 - 3) + 2\gamma(2\delta - 1) + 1]}\left\{\frac{M[2\gamma^2(\delta^2 - 1) + \gamma(2\delta - 1) + 1]}{\gamma + 1} - \frac{1}{M}\right\}$$

Further, using the appropriate relations, we can write the expression for the change in entropy across the shock front as

$$\frac{\Delta s}{\Gamma} = \frac{\gamma}{(\gamma-1)}\ln\left[\frac{\left[2+(\gamma-1)M^2\right]\left\{2\gamma\left[\gamma\left(\delta^2-1\right)+\delta\right]M^2-(\gamma-1)\right\}}{\left[4\gamma\delta\left(\gamma\delta-\gamma+1\right)+(\gamma+1)^2\right]M^2}\right]$$
$$-\ln\left[\frac{2\gamma\left\{\gamma\left(\delta^2-1\right)+\delta\right\}M^2}{(\gamma+1)}-\frac{(\gamma-1)}{(\gamma+1)}\right] \tag{6}$$

The forms of shock jump relations for non-ideal gas are similar to the well-known Rankine-Hugoniot conditions for an ideal gas [*see* Appendix] and the shock jump relations are explicitly written in terms of the upstream parameters only.

**3.1 Jump relations for weak shocks** For weak shock, $M$ is taken as $M = 1+\varepsilon$, where $\varepsilon$ is a parameter which is negligible in comparison with unity i.e., $\varepsilon << 1$. Thus, the pressure, density, particle velocity and sound speed just behind the weak shock can be, respectively, written as

$$p = \frac{p_o}{(\gamma+1)}\left\{\left[2\gamma^2\left(\delta^2-1\right)+\gamma(2\delta-1)+1\right]+4\gamma\left[\gamma\left(\delta^2-1\right)+\delta\right]\varepsilon\right\}$$
$$\rho = \frac{\rho_o\left[\gamma^2\left(4\delta^2-3\right)+2\gamma(2\delta-1)+1\right]}{(\gamma+1)^2}\left\{1+\frac{4\varepsilon}{(\gamma+1)}\right\} \tag{7}$$
$$u = \frac{4a_o}{\left[\gamma^2\left(4\delta^2-3\right)+2\gamma(2\delta-1)+1\right]}\left\{\gamma^2\left(\delta^2-1\right)+\gamma(\delta-1)+\left[\gamma^2\left(\delta^2-1\right)+(\gamma\delta+1)\right]\varepsilon\right\}$$
$$a \approx a_o$$

**3.2 Jump relations for strong shocks** For strong shock, $U >> a_o$, thus the pressure, density, particle velocity and sound speed just behind the strong shock can be, respectively, written as

$$p = \frac{2\rho_o\gamma\left[\gamma\left(\delta^2-1\right)+\delta\right]U^2}{\delta\gamma\left[2\gamma\delta^2-(\gamma-1)(2\delta-1)\right]}$$
$$\rho = \frac{\rho_o\left[\gamma^2\left(4\delta^2-3\right)+2\gamma(2\delta-1)+1\right]}{(\gamma+1)(\gamma-1)} \tag{8}$$
$$u = \frac{2\left[2\gamma^2\left(\delta^2-1\right)+\gamma(2\delta-1)+1\right]U}{\gamma^2\left(4\delta^2-3\right)+2\gamma(2\delta-1)+1}$$
$$a^2 = \frac{\gamma\delta(\gamma\delta-\gamma+1)U^2}{\gamma^2\left(4\delta^2-3\right)+2\gamma(2\delta-1)+1}$$

**3.3 Strength of a shock wave** In shock wave analysis, the quantity $\xi = (p-p_o)/p_o = p/p_o - 1$, represents the strength of the shock, and $\Delta p = p - p_o$ denotes the overpressure. The shock speed increases with increasing overpressure.

Substitution for $p/p_o$ from equation (8) yields

$$\xi = \frac{2\gamma\left[\gamma\left(\delta^2-1\right)+\delta\right]}{\gamma+1}M^2 - \frac{\gamma-1}{\gamma+1} - 1$$

$$\xi = \frac{2\gamma}{\gamma+1}\left\{\left[\gamma\left(\delta^2-1\right)+\delta\right]M^2-1\right\}$$

Thus, for shocks of any strength, we can write

$$\xi \propto \left\{\left[\gamma\left(\delta^2-1\right)+\delta\right]M^2-1\right\} \qquad \text{i.e.,} \quad \xi \propto \left(M^2-const.\right)$$

**Case I. For Shocks of vanishing strength** Shock waves for which $\xi$ is almost zero, are referred to as shocks of vanishing strength. For such shocks, $M=1$.

Thus for shocks of vanishing strength we can write $p/p_o \approx 1, \; \rho/\rho_o \approx 1, T/T_o \approx 1$ and $\Delta s/\Gamma \approx 0$.

**Case II. For Strong shocks** Since shock strength is proportional to $\left(M^2-const.\right)$, strong shocks are a result of very high values of upstream Mach number $M$. The maximum values of quantities in equations (4) and (5) are given below

$$\lim_{M\to\infty} M_{bs} = \sqrt{(\gamma-1)/2\gamma\,\delta\left(\gamma\,\delta-\gamma+1\right)}$$

$$\lim_{M\to\infty} p/p_o = \infty$$

$$\lim_{M\to\infty} \rho/\rho_o = \left\{\gamma^2\left(4\delta^2-3\right)+2\gamma\left(2\delta-1\right)+1\right\}/\left(\gamma^2-1\right)$$

$$\lim_{M\to\infty} T/T_o = \infty$$

## 4. Results and discussions

The generalized forms of jump relations for one-dimensional shock waves propagating in a non-ideal gas are derived which reduce to Rankine-Hugoniot shock conditions for ideal gas when non-idealness parameter becomes zero. The jump relations for pressure, density, temperature, and particle velocity are obtained, respectively in terms of upstream Mach number. The expression for change in entropy across the shock front is also obtained in terms of upstream Mach number. Further, the useful forms of shock jump relations for pressure, density and particle velocity in terms of non-idealness parameter are obtained for the two cases: viz., (i) when the shock is weak and (ii) when it is strong, simultaneously. The expression for downstream Mach number $M_{bs}$ is given by equation (4). The variations of downstream Mach number $M_{bs}$ with upstream Mach number $M$ for different values of non-idealness parameter $b\rho_o$ are shown through figure 1. It is notable that the downstream Mach number decreases with increase in the value of non-idealness parameter, as well as with the upstream Mach number.

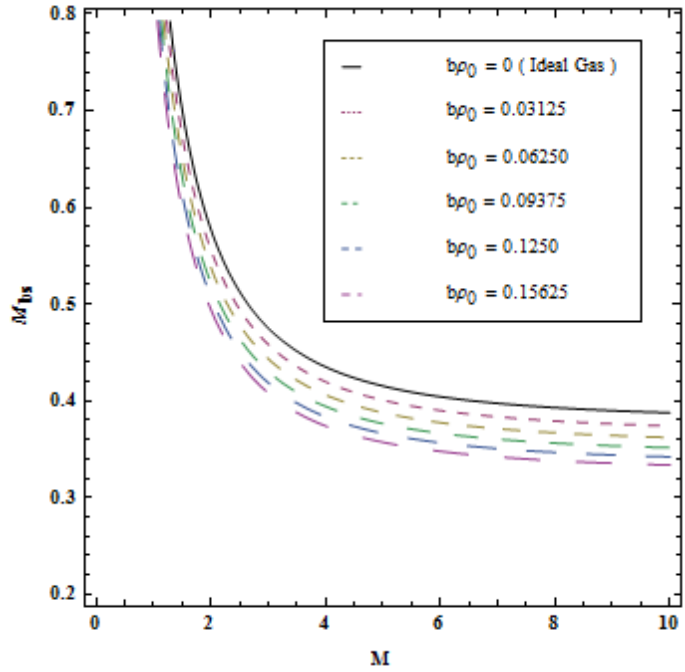

Fig. 1  The variation of $M_{bs}$ with $M$ for $\gamma = 1.4$ and different values of $b\rho_o$.

The generalized shock jump relations for the pressure $p / p_o$, density $\rho / \rho_o$, temperature $T / T_o$, and particle velocity $u / a_o$ are given by equation (5). The variations in the pressure, density, temperature, and particle velocity with upstream Mach number $M$, for $\gamma = 1.4$ and different values of non-idealness parameter $b\rho_o$, are shown in figure 2. It is important to note that the pressure, density, and particle velocity increase but the temperature decreases with increase in the value of non-idealness parameter. It is also seen that the pressure, density, temperatures and particle velocity increase with increase in the value of upstream Mach number. It is interesting to note that the increase in pressure and temperature can be infinitely large for sufficiently large shock strengths (or Mach number) but the density increase is limited to the value $\left\{\gamma^2\left(4\delta^2 - 3\right) + 2\gamma\left(2\delta - 1\right) + 1\right\}/\left(\gamma^2 - 1\right)$.

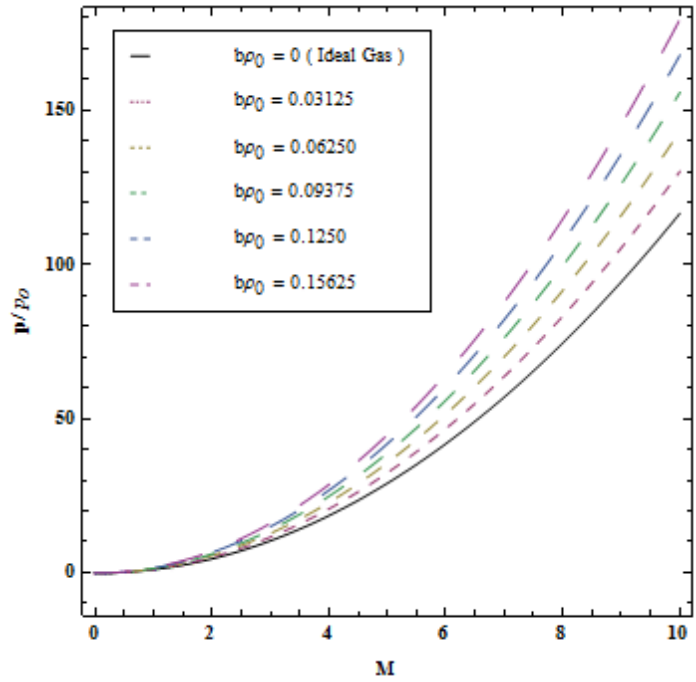

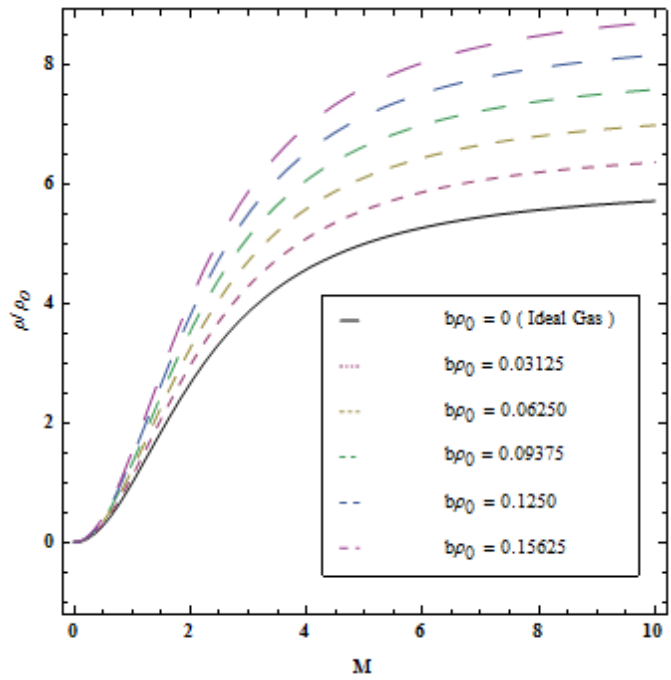

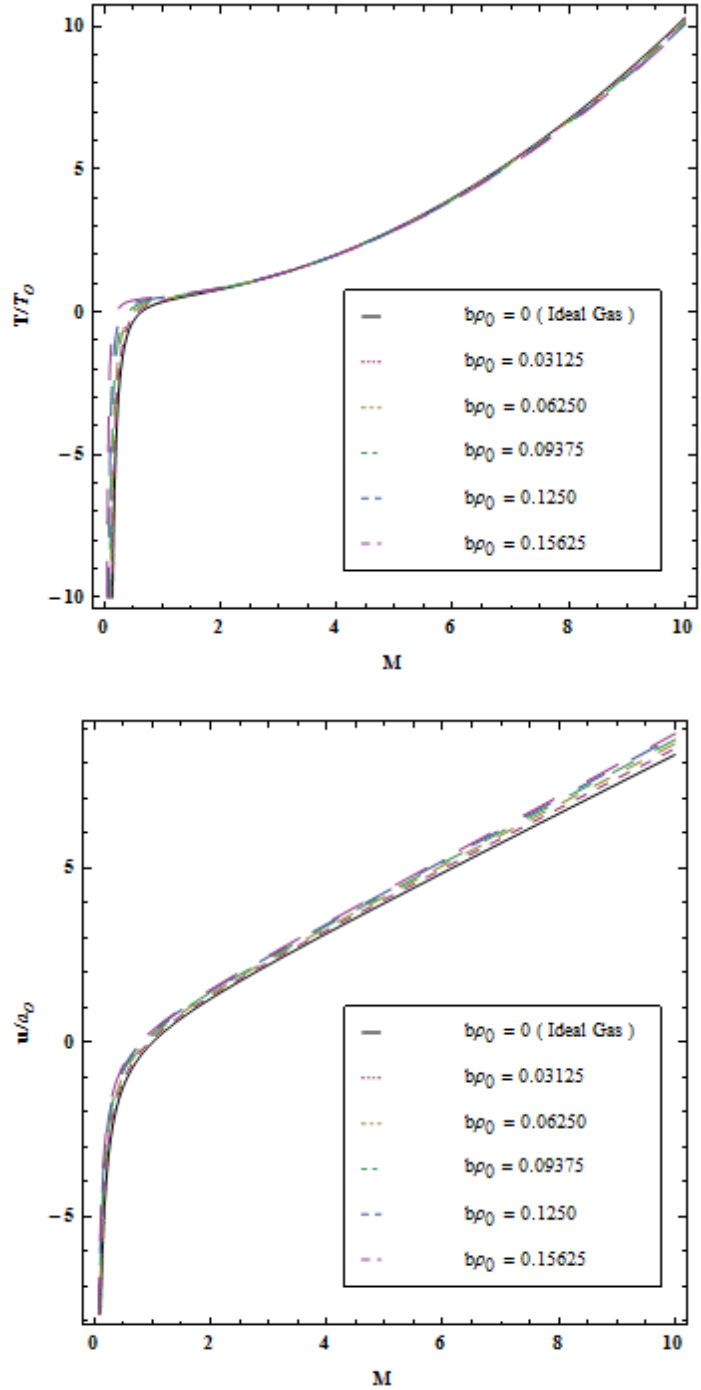

Fig. 2 The variations of $p/p_o$, $\rho/\rho_o$, $T/T_o$, and $u/a_o$ with $M$ for $\gamma = 1.4$ and different values of $b\rho_o$.

According to the second law of thermodynamics, the entropy of a substance cannot be decreased by internal processes alone, thus the downstream entropy in a shock must equal or exceed its upstream value, $s \geq s_o$. This entropy increase, predicted by the mass, momentum, and energy conservation relations alone, implies an irreversible dissipation of energy, even for an ideal fluid, entirely independently of the existence of a dissipation mechanism. For the sake of explanation equation (6), change in entropy $\Delta s/\Gamma$ is plotted with upstream Mach number $M$, for various values of non-idealness parameter $b\rho_o$ taking $\gamma = 1.4$ and shown through figure 3. Obviously, the change in entropy for ideal gas ($b\rho_o = 0$) is positive when upstream Mack number $M$ is greater than unity; hence, only a shock from supersonic to subsonic speed is possible, with a corresponding rise in pressure across the normal discontinuity. It is important to note that for curves except for $b\rho_o = 0$ (ideal gas) $\Delta s/\Gamma$ has always positive values for values of the upstream Mach number greater than 0.5 whereas $\Delta s/\Gamma$ has negative values for values of the upstream Mach number less than 0.5. Decrease in entropy is impossible by second law of thermodynamics, thus shock waves cannot develop in a flow where upstream Mach number is less than 0.5 in case of non-ideal gas or unity in case of ideal gas. It confirms that the shock waves may arise in a flow of non-ideal fluids where upstream Mach number is equal to or greater than 0.5 (approx).

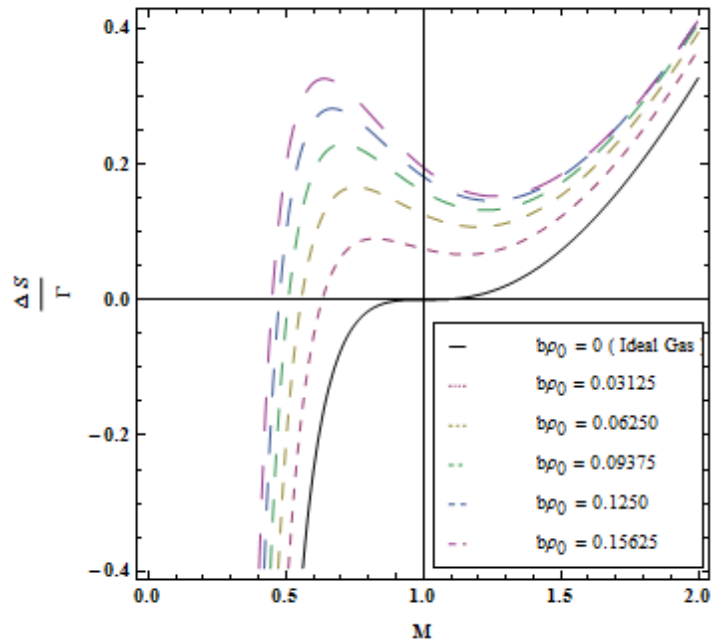

Fig. 3 The variations of $\Delta s/\Gamma$ with $M$ for different values of $b\rho_o$.

**4.1 Weak shock waves** The generalized jump relations for weak shock waves are given by equation (7). These relations are dependent of $\varepsilon(r)$ a parameter which is negligible in comparison with unity, adiabatic index $\gamma$ and non-idealness parameter $b\rho_o$. The variations in the pressure $p/p_o$, density $\rho/\rho_o$, and particle velocity $u/a_o$ with non-idealness parameter $b\rho_o$, for $\varepsilon = 0.2$, and different values of $\gamma$, are shown in figure 4. It is important to note that the pressure, density, and particle velocity increase with increase in non-idealness parameter. It is also seen that an increase in the value of $\gamma$ leads to an increase in the pressure whereas decrease in the particle velocity. The density first decreases up to $b\rho_o = 0.1$ and then starts to increase with increase in the value of $\gamma$.

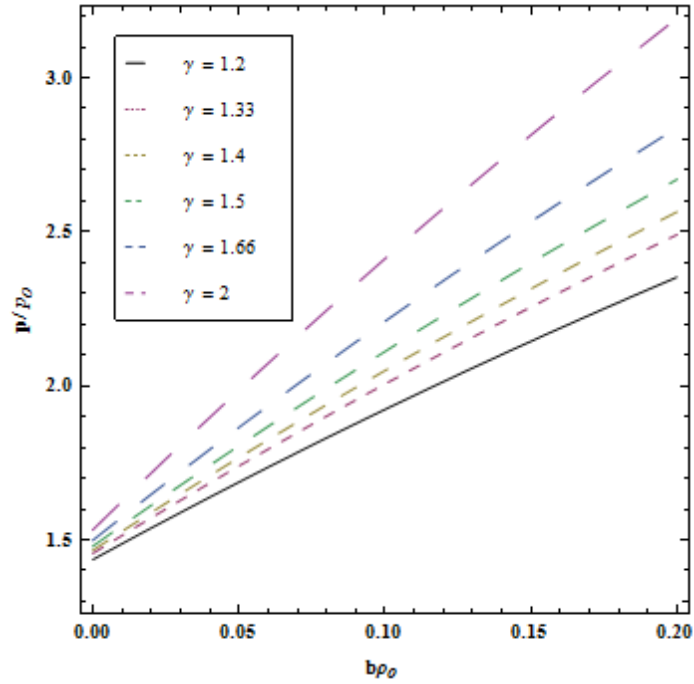

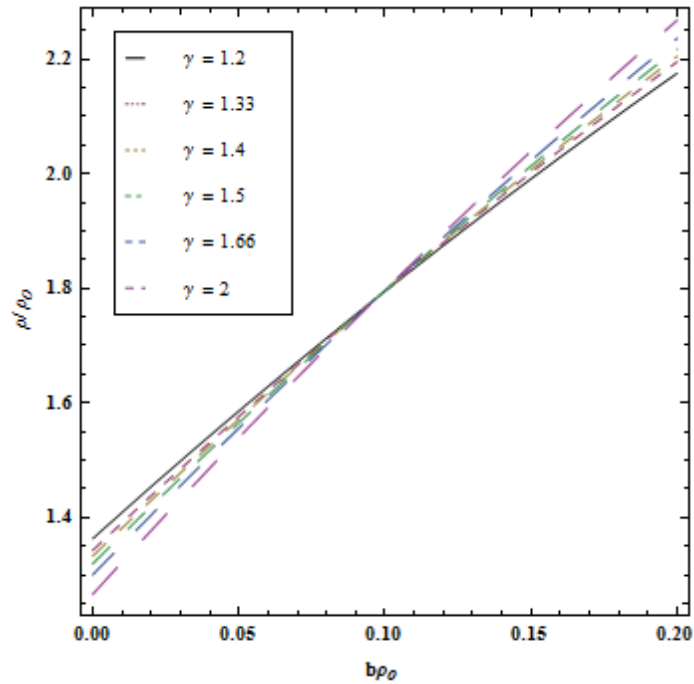

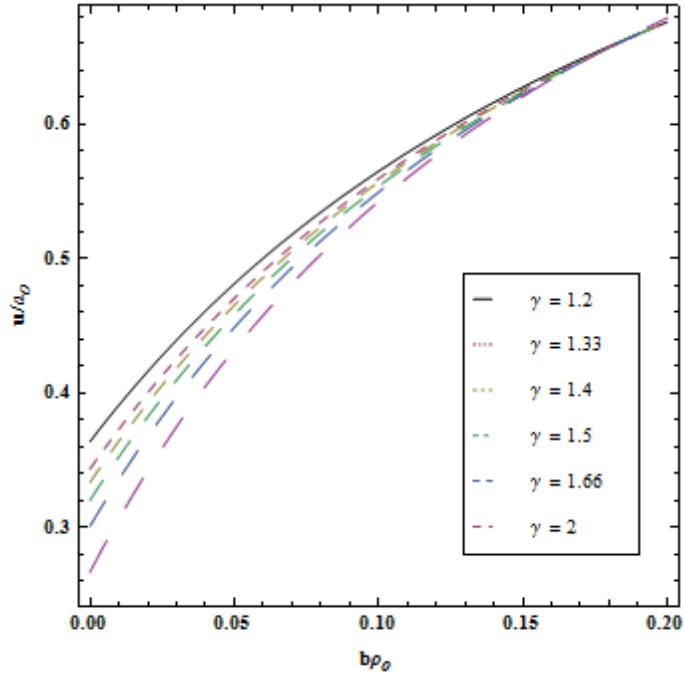

Fig. 4 The variations of $p/p_o$, $\rho/\rho_o$ and $u/a_o$ with $b\rho_o$ for $\varepsilon = 0.2$ and different values of $\gamma$.

**4.2 Strong shock waves** The generalized jump relations for strong shock waves are given by equation (8). These relations are dependent of the shock strength $U/a_o$, adiabatic index $\gamma$ and non-idealness parameter $b\rho_o$. The variations in the pressure $p/p_o$, density $\rho/\rho_o$, particle velocity $u/a_o$, and sound speed $a/a_o$ with non-idealness parameter $b\rho_o$ for $M = 20$ and different values of $\gamma$, are shown in figure 5. It is important to note that the pressure, density and particle velocity increase but the sound speed decreases with increase in non-idealness parameter. It is also seen that an increase in the value of $\gamma$ leads to an increase in the pressure whereas decrease in the density, particle velocity and sound speed.

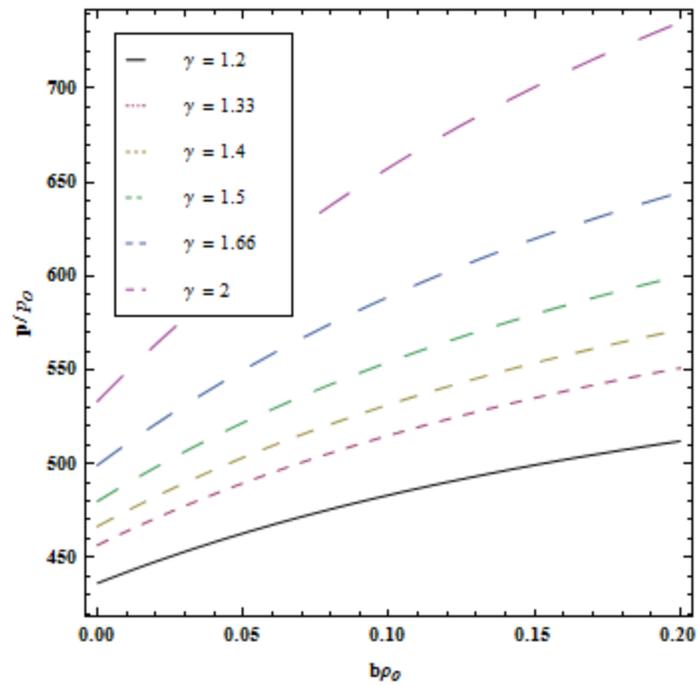

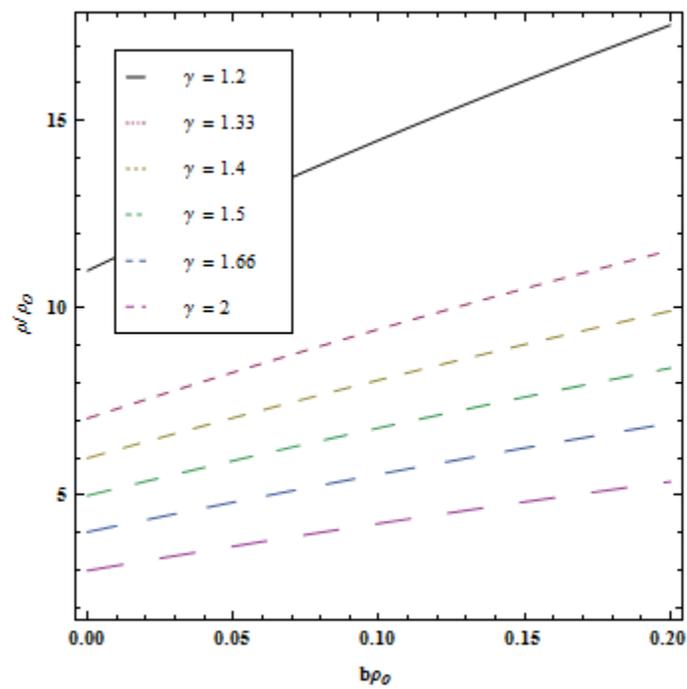

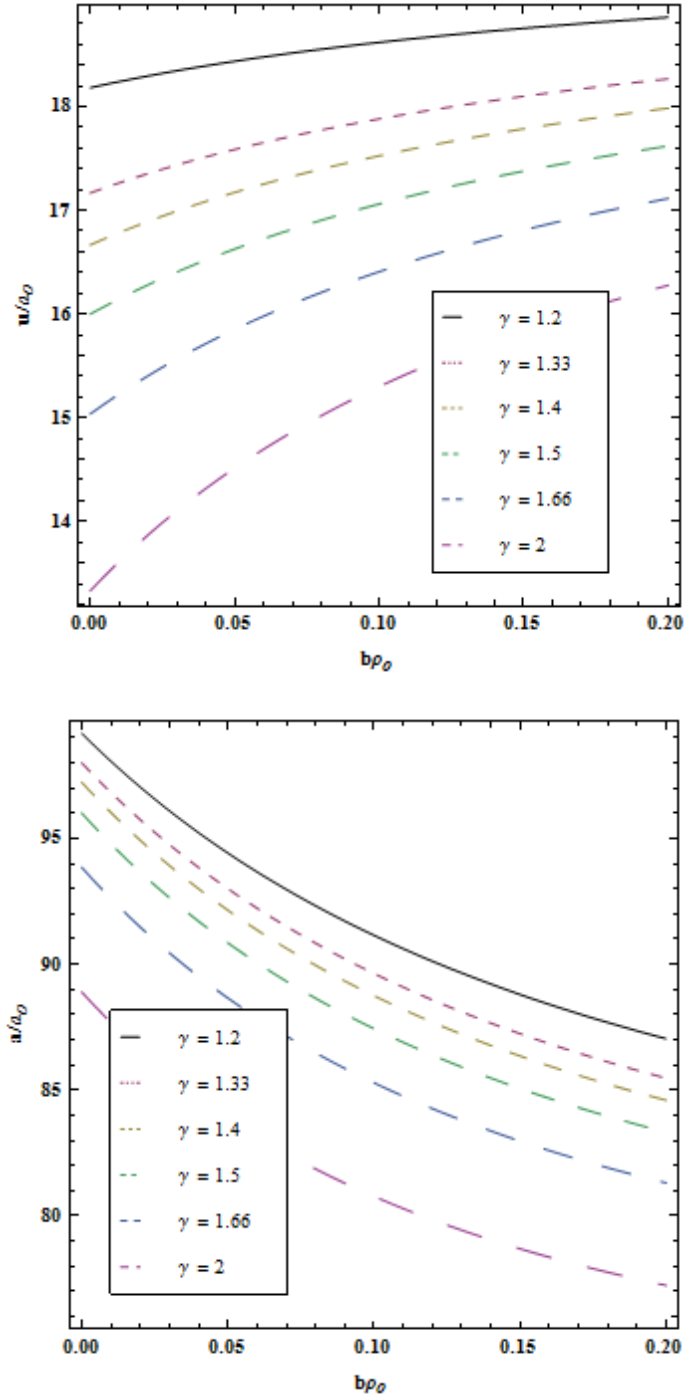

Fig. 5 The variations of $p/p_o, \rho/\rho_o, u/a_o$ and $a/a_o$ with $b\rho_o$ for $M=20$ and different values of $\gamma$.

**4.3 Applications** A study of the propagation of shock waves is essential for an understanding of the structure, evolution, and energy budget of the interstellar medium. The strongest shock waves in galaxies, apart from the violent phenomena in active galactic nuclei, are produced by supernova explosions. Investigations of supernovae remnants, which are the observational manifestations of supernova-shocks, give valuable information about the energetics of the supernova event as well as about the properties of the interstellar medium, where these shocks propagate. Shocks are thought to be important in acceleration of interstellar clouds. The interaction of blast waves with interstellar clouds plays an important role in the fundamental astrophysical process of star formation (Klein et al, 1994). The present analysis is aimed at a direct application of generalized jump relations (7) and (8) to the propagation of imploding shock waves through a non-ideal gas. The velocity of propagation of the shock depends on its strength. Flow variables (the pressure, density, entropy, and particle velocity) are all discontinuous over the shock front. The motion of a shock wave can be described by nonlinear equations of gas dynamics. According to the geometrical shock dynamics (Whitham, 1958, 1974), the characteristic form of the equations i.e., laws of conservation of mass, momentum and energy, governing the motion of a shock in non-ideal gas may be written as, $dp - \rho\, a\, du + \dfrac{\alpha\, \rho\, a^2 u}{u - a} \dfrac{dr}{r} = 0$. The geometrical factor $\alpha$ is to be taken as 0, 1 and 2 in the planer, cylindrical and spherical cases, respectively. On substituting the jump relations (7) or (8) into the characteristic equation a first order differential equation in $\varepsilon(r)$ or $U^2(r)$ is obtained which determines the shock.

### 4.3.1 Propagation of weak shock waves

Substituting the jump relations (7) into the characteristic equation, we get a first order differential equation in $\varepsilon(r)$ as

$$\left[\frac{4\delta_1}{\gamma\,\delta(\gamma+1)} - \frac{\delta_3}{\delta_2}\right]d\varepsilon + \left[\frac{4\delta_1}{\gamma\,\delta(\gamma+1)}\frac{dp_o}{p_o} - \frac{\delta_3}{\delta_2}\frac{da_o}{a_o} - \frac{\alpha\,\delta_3}{\delta_2}\frac{dr}{r}\right] = -\frac{(2\delta_1 - \gamma + 1)}{(\gamma+1)\gamma\,\delta}\frac{dp_o}{p_o}$$

where, $\delta_1 = \gamma\left[\gamma(\delta^2 - 1) + \delta\right]$, $\delta_2 = \gamma^2(4\delta^2 - 3) + 2\gamma(2\delta - 1) + 1$

and $\delta_3 = \gamma^2(4\delta^2 - 4) + 4(\gamma\delta + 1)$.

On integration, the differential equation yields $\varepsilon = K(\gamma\,\delta)^s r^{2\alpha s}$

where, $K$ is constant of integration and $s = \gamma\,\delta(\gamma+1)\delta_4\left[1 + \delta_4(\gamma+1)\gamma\,\delta/4\delta_1\delta_2\right]/8\delta_1\delta_2$.

Using the expression for $\varepsilon(r)$ and jump relations (7), one can easily determine the shock velocity, pressure, density and particle velocity immediately behind the weak shock front. The variations in the shock velocity $U/a_o$, pressure $p/p_o$, density $\rho/\rho_o$, and particle velocity $u/a_o$ with propagation distance $r$ for $\alpha = 2$, $\gamma = 1.4$ and different values of non-idealness parameter $b\rho_o$, are shown in figure 6. It is important to note that the shock velocity, pressure, density, and particle velocity increase for $b\rho_o = 0$, 0.03125, 0.06250 but remain unchanged for $b\rho_o = 0.09375$, 0.12500, 0.15625 with increase in propagation distance $r$.

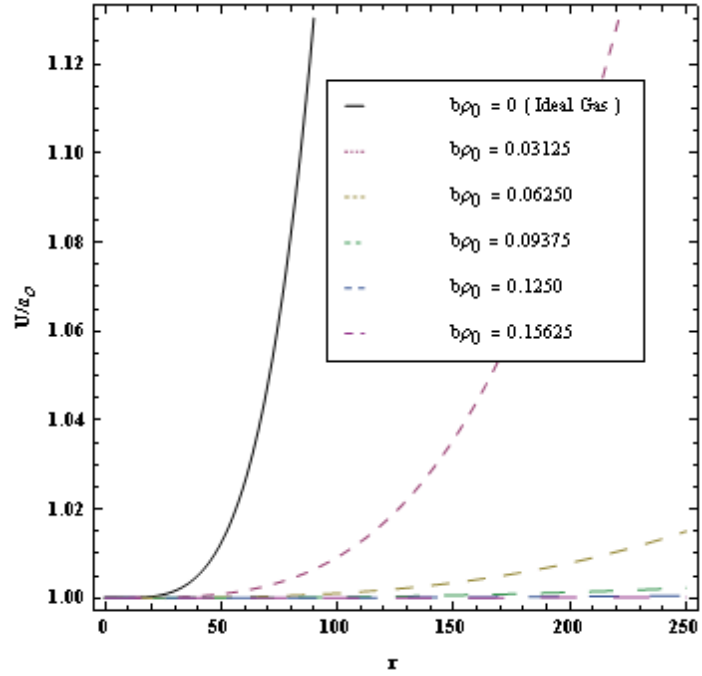

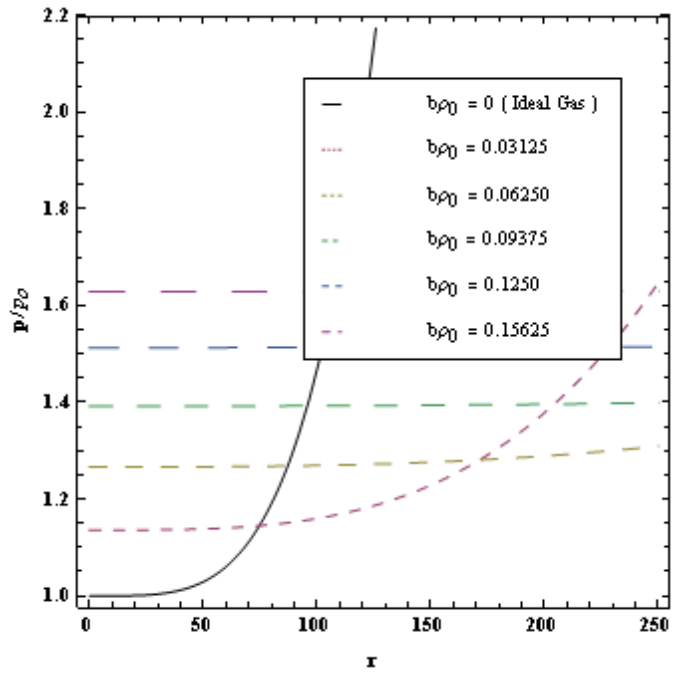

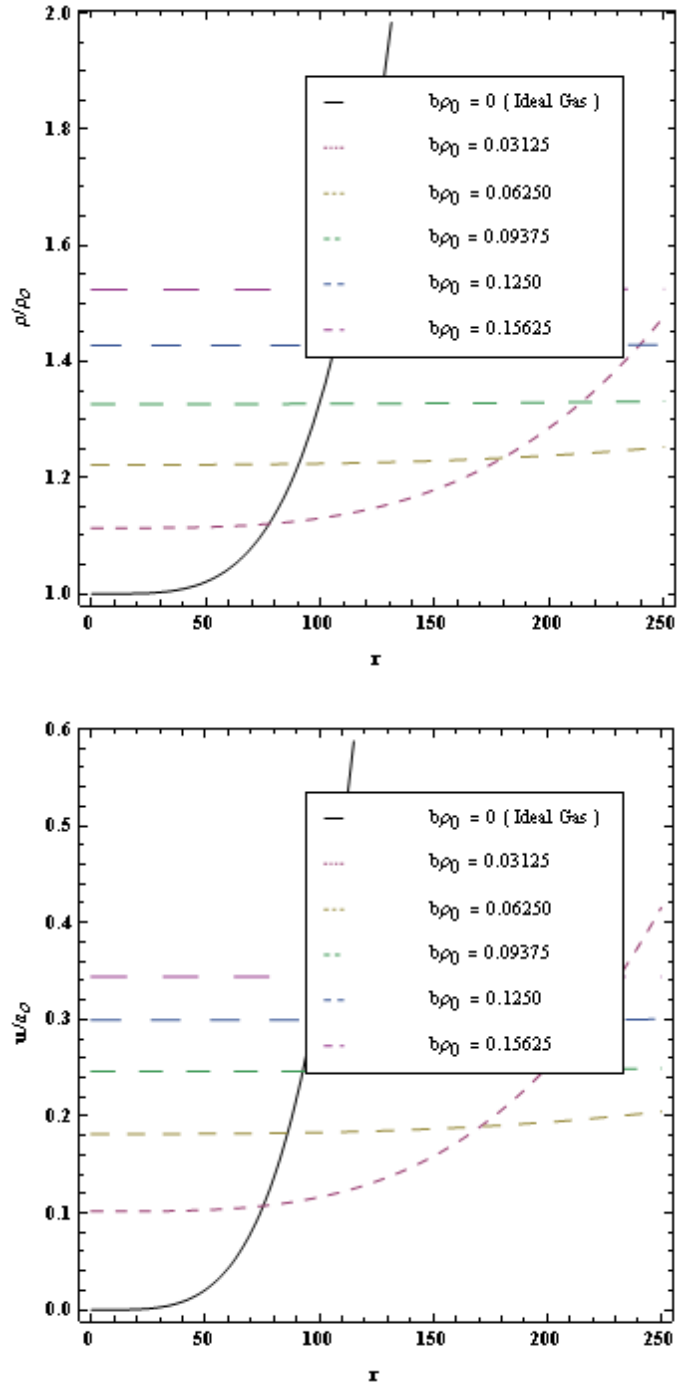

Fig. 6 The variations of $U/a_o$, $p/p_o$, $\rho/\rho_o$ and $u/a_o$ with $r$ for $\gamma = 1.4$, $\alpha = 2$ and different values of $b\rho_o$.

**4.3.2 Propagation of strong shock waves** Substituting the jump relations (8) into the characteristic equation, we get a first order differential equation in $U^2(r)$ as

$$\frac{dU^2}{dr} + \frac{2\alpha N U^2}{r} = 0$$

where $N = \dfrac{\delta_2 \delta_4 \delta_6}{2(\gamma^2 - 1)(\delta_4 - \delta_2\sqrt{\delta_6})} \left[ \dfrac{2\delta_1}{\delta_5} - \dfrac{\delta_4\sqrt{\delta_6}}{2(\gamma^2 - 1)} \right]^{-1}$, $\quad \delta_4 = \gamma^2(4\delta^2 - 4) + 2\gamma(2\delta - 1) + 2$,

$\delta_5 = \delta\gamma\left[ 2\gamma\delta^2 - (\gamma - 1)(2\delta - 1) \right]$ and $\delta_6 = \gamma\delta(\gamma\delta - \gamma + 1)/\delta_2$.

On integration, the differential equation yields

$$U = \sqrt{(K')} \, r^{-\alpha N}$$

where $K'$ is the constant of integration. This shows that the velocity of shock propagation is proportional to $r^{-\alpha N}$, where $r$ is the radius of the shock. Using the expression for $U(r)$ and jump relations (8), one can easily determine the shock velocity, pressure, density, particle velocity and sound speed immediately behind the strong shock front. The variations in the shock velocity $U/a_o$, pressure $p/p_o$, density $\rho/\rho_o$, particle velocity $u/a_o$, and sound speed $a/a_o$ with propagation distance $r$ for $\alpha = 2$, $\gamma = 1.4$ and different values of non-idealness parameter $b\rho_o$, are shown in figure 7. It is important to note that the shock velocity, pressure, particle velocity and sound speed decrease but the density remains unchanged with increase in the propagation distance. Obviously, the effect of non-idealness parameter $b\rho_o$ alters the numerical values of flow variables from their values for ideal gas case ($b\rho_o = 0$), but their trends of variations remain unchanged, in general.

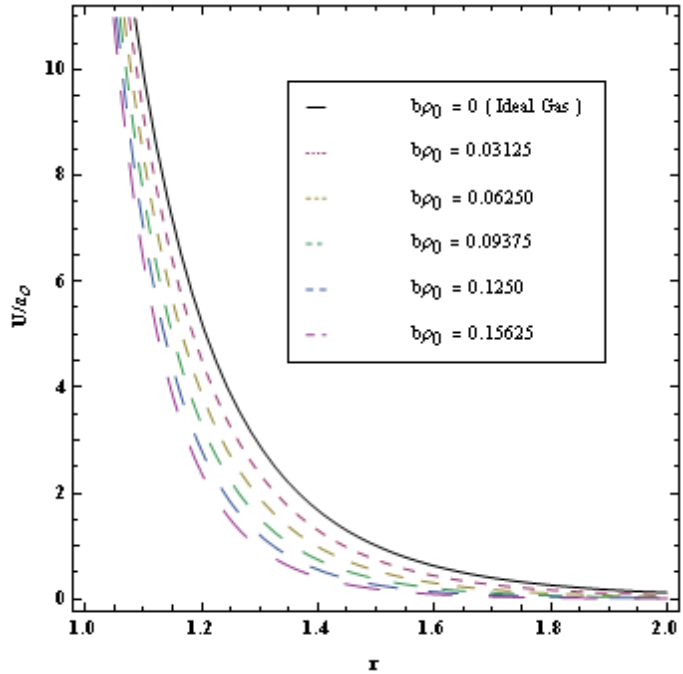

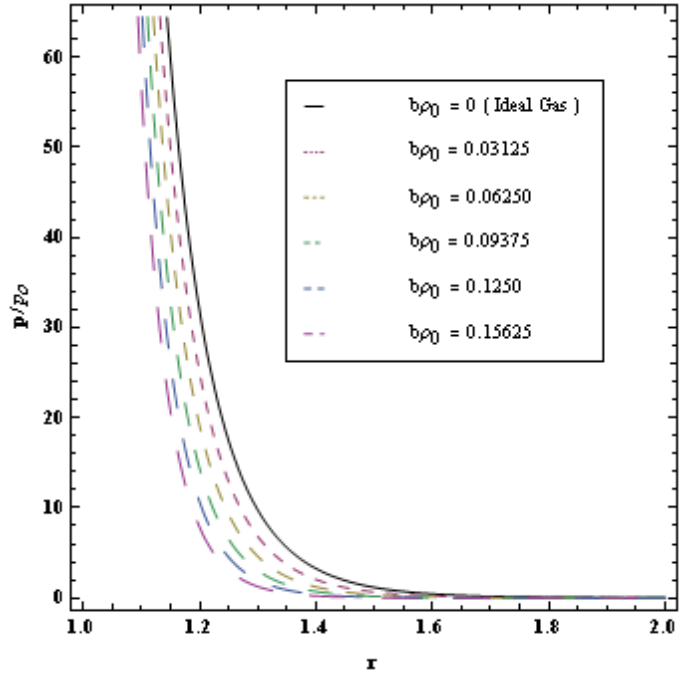

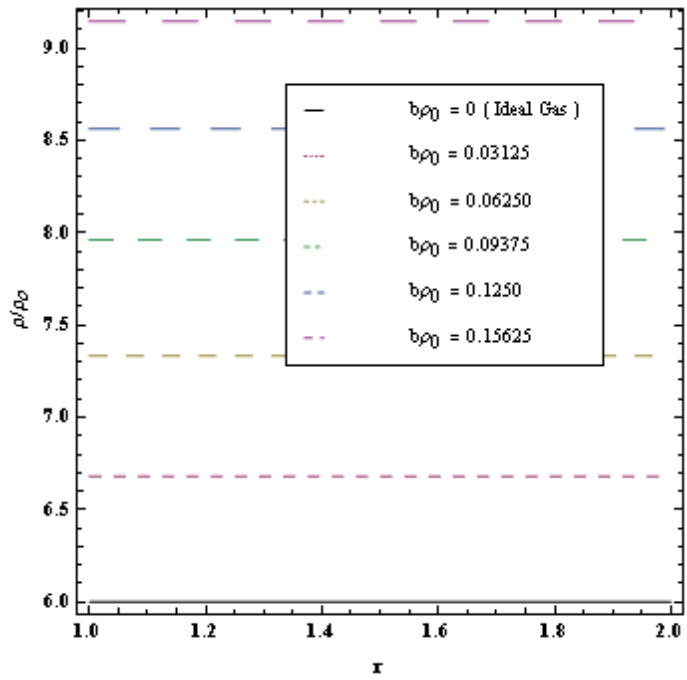

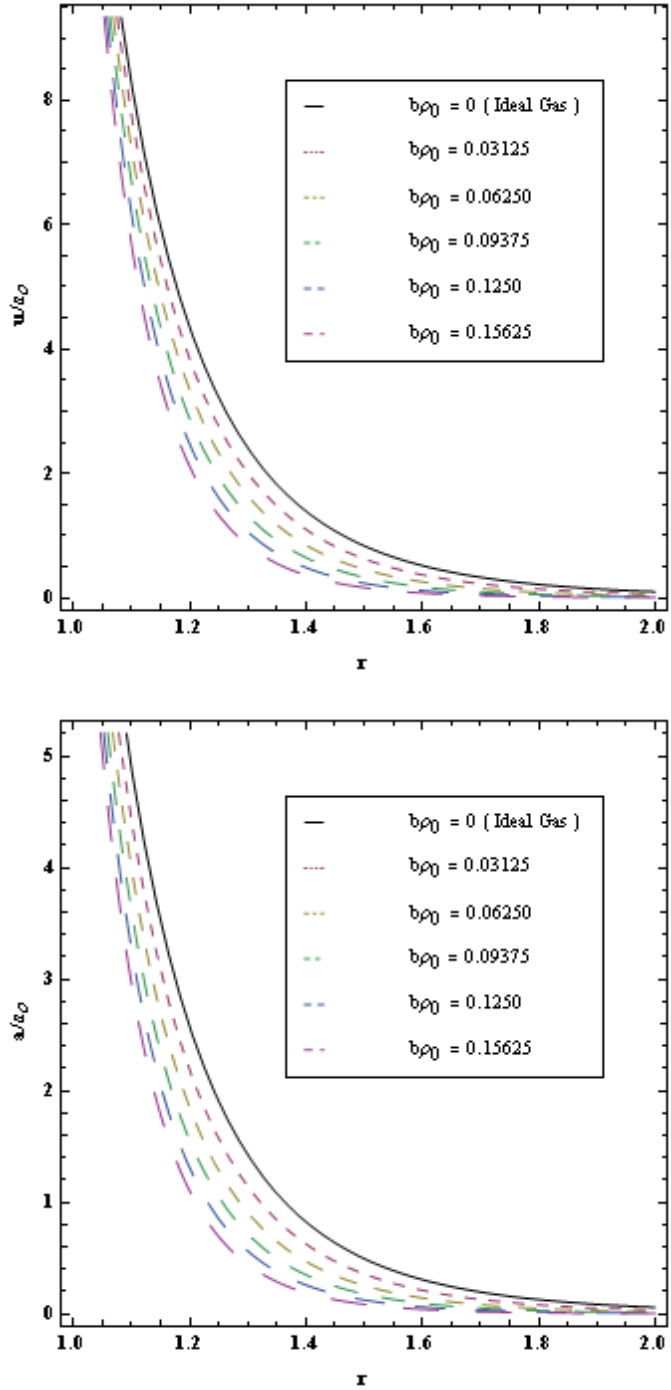

Fig. 7 The variations of $U/a_o$, $p/p_o$, $\rho/\rho_o$, $u/a_o$ and $a/a_o$ with $r$ for $\gamma = 1.4$, $\alpha = 2$ and different values of $b\rho_o$.

The entropy does not decrease across a shock front implies that $\Delta p > 0$, therefore rarefaction discontinuities do not exist. There are additional reasons why rarefaction discontinuities cannot exist. If such a discontinuity did exist, it would have $u_o < a_o$ and $u > a$ , and would therefore propagate subsonically through the unperturbed medium. But then any small disturbance, which would travel as an acoustic wave at the sound speed, produced in the flow at the jump could outrun the discontinuity. Therefore the rarefaction region behind the discontinuity would tend to spread into the gas in front of the discontinuity faster than the discontinuity itself could propagate, and in doing so would erode away any initial jump in material properties. This means that a rarefaction discontinuity immediately smoothes into a continuous transition. Furthermore, because a rarefaction discontinuity would move supersonically with respect to the downstream material, it could not be affected by any process or change in conditions occurring behind the jump. That is, no feedback on the wave is possible, and in that sense the wave is uncontrolled. Both of these properties imply that the rarefaction discontinuities are mechanically unstable, and disintegrate immediately.

In contrast, the entropy increases in a compression shock. The front outruns acoustic waves that might tend to smear it out, and the upstream material remains "unaffected" of the shock until it slams into it; hence the shock can propagate as a sharp discontinuity. Furthermore, the shock propagates subsonically with respect to the downstream material, hence the material behind the front can affect the front's behavior; if the downstream gas is strongly compressed and heated, it tends to strengthen the shock; if the downstream material cools rapidly (e.g., by radiation losses) the driving force behind the shock front weakens and finally the shock dissipates. Thus, the compression shocks satisfy entropy constraints and are mechanically stable.

## 5. Conclusions

The present article deals with the formulation of generalized shock jump relations between the upstream and downstream quantities. The useful forms of jump relations for weak and strong shocks provide simpler and more accurate boundary conditions to determine the shock structure and propagation of shocks in non-ideal gases. However, although not exact, these equations have two advantages: (i) all shock relations are explicitly written in terms of the upstream parameters only, and (ii) the forms of the expressions are similar to the well-known Rankine-Hugoniot relations for ideal gas. The results of this work confirm that the shock waves may arise in the flow of non-ideal fluids where upstream Mach number is less than unity i.e., $M \leq 0.5$ (approx).

A Shock always moves supersonically relative to the gas ahead. This means that no information about the flow disturbance can reach the gas ahead of the shock before the gas enters the shock. But the shock travels subsonically relative to the gas behind it, so that disturbances arising in the downstream gas will generate waves which will finally overtake the shock and modify its strength. The strength of a shock can increase or decrease in time under the influence of several processes. For one, dissipation of the shock energy into heat through viscosity and thermal conduction tends to reduce the shock strength. The rate of this dissipation is of central importance in the calculations of shock heating of the chromospheres and corona. Shocks propagating in more than one dimension are weakened simply because the original impulse is diluted over a larger surface area. A shock can also increase or decrease its strength by propagation into media of different temperature and pressure. Important effects of this kind arise in the solar atmosphere, where outward movement of shocks into increasingly rarefied plasma above the photosphere tends to increase their strength to satisfy the conservation laws given by equation (3). They weaken, however, when they begin to propagate above the temperature minimum into the rapidly increasing temperatures of the chromospheres, where the sound speed increases outward, and the disturbance tends to be 'stretched out'(the inverse of the steepening), thus decreasing its peak amplitude.

It is also desirable to mention that the effect of non-idealness parameter, in general, does not change the trends of variations of flow variables behind the weak and strong shocks. In my view, the shock wave problems can be generalized by applying the generalized jump relations and the results obtained can be easily reduced to the case of shocks in an ideal gas atmosphere. Obviously, the new results will be closer to the actual situations. Thus, the generalized jump relations have a key role in the research of shocks in air, stars, interstellar medium, etc.

## Appendix

Shock jump relations for ideal gas (Whitham, 1958)

$$p = \rho_o \, a_o^2 \left\{ \frac{2M^2}{\gamma+1} - \frac{\gamma-1}{\gamma(\gamma+1)} \right\}$$

$$\rho = \rho_o \frac{(\gamma+1)M^2}{(\gamma-1)M^2+2}$$

$$T = \frac{T_o \left[2+(\gamma-1)M^2\right]\left\{2\gamma \, M^2 - (\gamma-1)\right\}}{(\gamma+1)^2 \, M^2}$$

$$u = a_o \frac{2}{\gamma+1}\left(M - \frac{1}{M}\right)$$

$$\frac{\Delta s}{R} = \frac{\gamma}{(\gamma-1)}\ln\left[\frac{2+(\gamma-1)M^2}{(\gamma+1)M^2}\right] + \frac{1}{\gamma-1}\ln\left[\frac{2\gamma M^2}{(\gamma+1)} - \frac{(\gamma-1)}{(\gamma+1)}\right]$$

where $M = U/a_o$ and $a_o^2 = \gamma \, p_o/\rho_o$

Jump relations for weak shocks in ideal gas (Anand, 2000)

$$p = p_o\left\{1 + \frac{4\gamma\varepsilon}{\gamma+1}\right\}, \rho = \rho_o\left\{1 + \frac{4\varepsilon}{\gamma+1}\right\}, \; T = T_o\left\{1 + \frac{8\gamma(\gamma-1)\varepsilon}{(\gamma+1)^2}\right\},$$

$$u = \frac{4a_o}{\gamma+1}\varepsilon \; \text{and} \; U = a_o(1+\varepsilon)$$

Jump relations for strong shocks in ideal gas (Yousaf, 1987)

$$p = \frac{2}{\gamma+1}\rho_o U^2, \; \rho = \frac{(\gamma+1)}{(\gamma-1)}\rho_o, \; T = 2T_o\frac{(\gamma-1)\gamma}{(\gamma+1)^2}\left(\frac{U}{a_o}\right)^2,$$

$$u = \frac{2}{\gamma+1}U \; \text{and} \; a = \frac{\sqrt{2\gamma(\gamma-1)}}{\gamma+1}U$$